\documentclass{article}

\usepackage{graphicx}  % standard LaTeX graphics tool;
\usepackage{amsmath}   % For getting proper math eqns
\usepackage[compress]{cite}
\usepackage{amssymb}   % Used for getting various symbols eg. gtrsim, lesssim etc.
\usepackage{bm} % For bold math (esp of lower greek letters)

\usepackage{mathrsfs}

\def\eq#1{{Eq.~(\ref{#1})}}

\title{Demystifying the constancy of the Ermakov-Lewis invariant for a time dependent oscillator}

\author{T. Padmanabhan\\
IUCAA, Post Bag 4, Ganeshkhind,
 Pune - 411 007, India.\\
email: paddy@iucaa.in}

\date{ }

\begin{document}

\maketitle

\begin{abstract}
It is well known that the time dependent harmonic oscillator possesses a conserved quantity, usually called Ermakov-Lewis invariant. I provide a simple physical interpretation of this invariant as well as   a whole family of related invariants. This interpretation  does not seem to have been noticed in the literature before. The procedure also allows one to tackle some key conceptual issues which arise in the study of quantum fields in external, time dependent, backgrounds like in the case of particle production in an expanding universe and Schwinger effect. 
\end{abstract}

\section{Introduction}

The time dependent harmonic oscillator (TDHO)  arises in several physical situations and has been studied extensively  in the past. There are several methods to study TDHO, of which the one based on Ermakov-Lewis invariant is quite popular (see, for e.g., Ref.~\cite{ref1}; for  a bit of history and  references, see Ref.~\cite{ref2}). This invariant can be introduced as follows. Let $q(t)$ satisfy the \textit{linear} harmonic oscillator equation with a time dependent\footnote{Usually, in the literature, one works with $m=1$ since this can always be achieved by a change of the independent variable. We shall keep $m(t)$ time dependent for the sake of generality.} mass $m(t)$ and frequency $\omega(t)$:
\begin{equation} 
\frac{d}{dt} (m \dot q)  + m \omega^2 q =0
 \label{tp12}
\end{equation} 
where the overdot denotes differentiation with respect to $t$.  Let $f(t)$ be another function which satisfies the \textit{non-linear} differential equation: 
\begin{equation}
 \frac{d}{dt}(m\dot f) + m \omega^2 f =\frac{\Omega^2}{mf^3}
 \label{tp24}
\end{equation}
where $\Omega$ is a real constant.\footnote{This constant is usually  set to unity in the literature but, as we shall see, it is useful to keep it as a free parameter.}
It is then straightforward to verify that the following quantity 
\begin{equation}
 I= \frac{1}{2} \left[ m^2 ( \dot q f - q \dot f )^2 + \frac{\Omega^2 q^2}{f^2}\right]
 \label{tp45}
\end{equation} 
is conserved. That is, $dI/dt=0$. (The $I$, with $m=1, \Omega=1$ is called Ermakov-Lewis invariant.)
This is the simplest of a family of invariants one can introduce in the study of TDHO. 

The question arises as to whether $I$ has any physical significance and whether its invariance can be understood in a simple manner. I will show that this is indeed possible. One can eliminate the time dependence of the oscillator variable $q(t)$ by a suitable transformation and reduce the system to the study of another oscillator, $Q(t)$, with constant frequency $\Omega$ and unit mass. The energy  of the $Q$ oscillator, which of course is conserved, turns out to be the Ermakov-Lewis invariant. I was led to this interpretation of Ermakov-Lewis interpretation while studying QFT in an external field and was motivated to write this note because this approach does not seem to have been \textit{explicitly} mentioned in the rather extensive literature (though it is might probably be hidden implicitly in some of the earlier work), to the best of my knowledge. Moreover, by a slight generalization of this approach, one can discover a family of invariants obtained earlier (without any physical interpretation) in Ref.~\cite{ref3}. I will now show how all these can be done, in a fairly straightforward and simple manner. 

\section{The Ermakov-Lewis invariant is just the energy}

Start with the Lagrangian describing the original TDHO variable $q(t)$ given by 
\begin{equation}
L_q = \frac{1}{2} m (t) \left[ \dot q^2  - \omega^2(t)q^2\right]
\label{tp1}
\end{equation} 
and introduce a new dynamical variable $Q$ and a new time coordinate $\tau$ through the transformations 
\begin{equation}
 Q \equiv \frac{q}{f(t)}; \qquad dt \equiv mf^2 d\tau
 \label{tp2}
\end{equation}
where $f(t)$ satisfies \eq{tp24}.
A simple calculation shows that the Lagrangian in \eq{tp1} now becomes 
\begin{equation}
L_Q = \frac{1}{2}\left[ Q'^2 - \Omega^2 Q^2 \right] + \frac{d}{d\tau} \left( \frac{1}{2} \frac{f'}{f} Q^2 \right)
 \label{tp3}
\end{equation} 
where the prime denotes derivative with respect to $\tau$ and $\Omega^2$ is a new frequency which will be a constant if $f$ satisfies \eq{tp24}. Since the total time derivative in the Lagrangian $L_Q$ does not affect the equations of motion, the $Q$-oscillator will satisfy\footnote{One can  directly verify that \eq{tp12} will acquire this form under the transformations in \eq{tp2}, though it is obvious from the form of the Lagrangian.} the equation $Q'' +\Omega^2 Q =0$.  For this conservative $Q$-system, the energy will be a constant and it is given by 
\begin{equation}
E = \frac{1}{2} (Q'^2 + \Omega^2 Q^2) = \frac{1}{2} \left[ m^2( \dot q f - q \dot f )^2 + \frac{\Omega^2 q^2}{f^2}\right]
 \label{tp46}
\end{equation} 
This is precisely the Ermakov-Lewis invariant discussed so extensively in the literature; it is just the energy of the $Q$-oscillator!

In fact, this approach has a very natural generalization which allows us to find invariants for different kinds of non-linear differential equations. Let me briefly illustrate one possible generalization which which will allow us to provide an interpretation for a result which has appeared in the literature\cite{ref3} before. Consider a dynamical system with a Lagrangian $L_Q = (1/2) Q'^2 - V(Q)$ and the conserved energy $E=  (1/2) Q'^2 + V(Q)$. (The system we studied so far corresponds to $V(Q) = (1/2)\Omega^2 Q^2$. We are now generalizing it to an arbitrary potential.)
This energy $E$ will, of course, be an invariant in terms of any other set of variables.
Introduce a new dynamical variable $q=fQ$ and a new time variable $\tau$ with $mf^2 d\tau = dt$, keeping $f(t)$  an arbitrary function of time, at this stage. 
This will transform the Lagrangian $L_Q = (1/2) Q'^2 - V(Q)$ to the form
\begin{equation}
L_q = \frac{1}{2} m\dot q^2 + \frac{1}{2}\frac{q^2}{f} \frac{d}{dt} (m \dot f) - \frac{V(q/f)}{mf^2}
\end{equation} 
where dot denotes derivative with respect to $t$ and we have ignored a total time derivative. The dynamical variable $q$ will satisfy the equations of motion arising from this Lagrangian: 
\begin{equation}
 \frac{d}{dt} (m \dot q) = \frac{q}{f}\frac{d}{dt} (m \dot f)  - \frac{V'}{mf^3}
\end{equation} 
where $V'$ denotes derivative of $V$ with respect to its argument. We now demand that this equation should be identical to \eq{tp12}, which, in turn, determines $f$ to be a solution of the equation
\begin{equation}
 \frac{d}{dt} (m \dot f)  + m \omega^2 f  =\frac{F(q/f)}{mf^3}
 \label{fnew}
\end{equation} 
where $F(Q) = V'(Q)/Q$ is a function of $(q/f)$. The conserved energy of the original system can now be expressed in a form which is precisely the class of invariants found in Ref.\cite{ref3} but expressed in our notation:
\begin{equation}
 E =  \frac{1}{2} Q'^2 + V =  \frac{1}{2} m^2 ( \dot q f - q \dot f)^2 + V =  \frac{1}{2} m^2( \dot q f - q \dot f)^2 + \int dq \ \frac{F}{f^2} q
\label{enew}
 \end{equation} 
This result can be summarized as follows:
 Consider a pair of differential equations for the two variables $f(t),q(t)$ where $f(t)$ satisfies \eq{fnew} --- which contains an arbitrary function $F(q/f)$ --- and $q(t)$ satisfies \eq{tp12}. Such a system has a conserved quantity given by
\eq{enew} where $V$ and $F$ are related by $F(Q) = V'(Q)/Q$. This result  shows the power of the Lagrangian based transformation to discover the invariants to differential equations. 

\section{Discussion}

The trick of transforming the TDHO to a time \textit{independent} oscillator with constant frequency has several conceptual and  practical applications which I will briefly mention here. (These applications will be discussed in detail elsewhere.) 

Several problems in quantum field theory in external fields can be reduced to that of TDHO. Two prominent examples are: (i) particle production in an expanding cosmological background and (ii) particle production in a time dependent electric field (Schwinger effect). In both cases, expanding the field in terms of Fourier modes 
$\exp(i\mathbf{k}\cdot \mathbf{x}$) reduces the problem to that of TDHO for each mode labeled by $\mathbf{k}$. For example, the study of a massless scalar field in the FRW universe can be reduced to that of a TDHO with $m(t)=a^3(t)$ and $\omega(t)=k/a(t)$ where $a(t)$ is the expansion factor and $t$ is the cosmic time. Similarly the QFT of a charged scalar field in a external time-dependent electric field $E(t)$ can be reduced to the study of a TDHO with unit mass and $\omega(t)$ determined by $E(t)$.

One conceptual difficulty encountered in the study of these systems is in defining a natural vacuum state and the notion of a particle in the presence of time-dependent background.
Our approach provides a simple solution to this problem. Once the TDHO is transformed to a time \textit{independent} oscillator, we obtain a preferred choice of ground state for $Q$ which can be translated to the $q$-oscillator. It is straightforward to relate the quantum dynamics of $Q$-oscillator to that of the original $q$-oscillator by elementary algebra. For example, if $\phi(t,Q)$ is a solution to the Schrodinger equation for the $Q$-oscillator, then the corresponding solution $\psi(t,q)$ to the Schrodinger equation satisfied by the $q$-oscillator is given by:
\begin{equation}
\psi(t,q) = \frac{1}{\sqrt{f}}\ \phi \left[ \tau(t) , Q=(q/f)\right] \, \exp\left(i \frac{m\dot f}{2f} q^2\right)
 \label{tp10}
\end{equation} 
The $(1/\sqrt{f}) $ factor takes care of the proper normalization of the wave function while the extra phase factor arises from the total time derivative term in \eq{tp3}. This result completely solves the problem of translating the wave function from one system to another in the Schroedinger picture. 

In the Heisenberg picture, we are usually interested in the Bogoliubov coefficients relating the creation and annihilation operators $(A,A^\dagger)$ of the $Q$ oscillator to those $(a,a^\dagger)$ of $q$ oscillators. The relevant  Bogoliubov transformation is given by
$A=\alpha(t) a + \beta(t) a^\dagger$ where:
\begin{equation}
 \alpha = \frac{1}{2} \left(\frac{m}{\omega\Omega}\right)^{1/2} \, f \left[ \omega + \frac{\Omega}{mf^2} - i \frac{\dot f}{f}\right] 
 \label{tp17}
\end{equation} 
with time dependent $\omega(t), m(t)$  and
\begin{equation}
 \beta =-\frac{1}{2} \left(\frac{m}{\omega\Omega}\right)^{1/2} \, f \left[ \omega - \frac{\Omega}{mf^2} + i \frac{\dot f}{f}\right] 
\label{tp18}
\end{equation} 
If we set the initial frequency $\omega(t_i)=\Omega$, then these  Bogoliubov coefficients completely determine the future evolution. All other quantities, like path integral kernel, Green functions, in-out amplitude etc can all be computed in a similar fashion.
Thus the procedure allows us to define a  preferred vacuum state in an external time dependent background and calculate all the physical quantities in a simple manner. 

\textit{Acknowledgements:} The research work of the author is partially supported by the J.C. Bose research grant of DST, India.

\end{document}